\documentclass{elsart}
\usepackage[latin1]{inputenc}
\usepackage{amssymb}
\date{}

\newtheorem{The}{Theorem}[section]

\newtheorem{Deff}[The]{Definition}
\newtheorem{Lem}[The]{Lemma}
\newtheorem{Rem}[The]{Remark}
\newtheorem{Exa}[The]{Example}
\newtheorem{Cla}[The]{Claim}

\newcommand{\fa}{\forall}
\newcommand{\Ga}{\Gamma}

\newcommand{\Si}{\Sigma}
\newcommand{\Sio}{\Sigma^\omega}
\newcommand{\Sis}{\Sigma^\star}
\newcommand{\ra}{\rightarrow}
\newcommand{\hs}{\hspace{12mm}

\noi}

\newcommand{\Exab}{\begin{Exa}}
\newcommand{\Exae}{\end{Exa}}

\newcommand{\lra}{\leftrightarrow}
\newcommand{\la}{language}
\newcommand{\ite}{\item}
\newcommand{\vp}{\varphi}
\newcommand{\Lp}{L(\varphi)}

\newcommand{\loc}{local language}
\newcommand{\locs}{local sentence }

\newcommand{\ol}{ $\omega$-language}
\newcommand{\orl}{ regular $\omega$-language}
\newcommand{\om}{\omega}
\newcommand{\nl}{\newline}
\newcommand{\noi}{\noindent}
\newcommand{\proo}{\noi {\bf Proof.} }
\newcommand {\ep}{\hfill $\square$}

\begin{document}
\begin{frontmatter}

\title{\bf CLOSURE PROPERTIES OF \\ LOCALLY FINITE
$\omega$-LANGUAGES\thanksref{titlefn}}
\thanks[titlefn]{Partially supported by Intas 00-447.}

\author{Olivier Finkel\corauthref{cor}}
\corauth[cor]{Corresponding author}
\ead{finkel@logique.jussieu.fr }

\address{ Equipe de Logique Math\'ematique \\CNRS  et
Universit\'e Paris 7,  U.F.R. de Math\'ematiques \\  2 Place Jussieu 75251
Paris
 cedex 05, France.}

\vspace{10mm}
\hfill
\small{\it \bf  Dedicated to Denis Richard for his 60 th Birthday}

\begin{abstract} \noi Locally finite $\omega$-languages,
defined via second order quantifications followed by a
first order locally finite sentence,  were introduced by
Ressayre in \cite{ress}. They enjoy  very nice properties  and
extend \ol s accepted by finite automata or defined
by monadic second order sentences.
We study here closure properties of the family $LOC_\om$ of
 locally finite omega languages. In particular we show that the class
$LOC_\om$
is neither closed under intersection nor under
complementation, giving an answer to a question of Ressayre \cite{ress2}.
\end{abstract}

\end{frontmatter}

\section{Introduction}

In the sixties J.R. B\"uchi was the first to study \ol s recognized by
finite
automata in order to prove the decidability of the monadic second order
theory of one
successor over the integers \cite{bu62}. In the course of his proof he
showed that an \ol ,
i.e. a set of infinite words  over a finite alphabet,
is accepted by a finite automaton with the now called B\"uchi acceptance
condition
if and only if it is defined by an (existential)
monadic second order sentence. Algorithms have been
found to give such an automaton from the monadic second order sentence and
conversely. Thus  the  above cited decision problem is reduced to the
decidability
of the emptiness problem for B\"uchi automata which is easily shown to be
decidable.
 The equivalence between definability by monadic second order sentences
and acceptance by finite automata holds also  for  languages
 of finite words \cite{bu60}, and has been extended  to
 languages of words of length $\alpha$, where $\alpha$ is a
countable ordinal $\geq \om$ \cite{bs}.

 The research area, now called ``descriptive complexity",  found its origin
in the above
cited work of B\"uchi as well as in the fundamental result of Fagin who
proved  that the class
{\bf NP} is characterized by existential
second order formulas, \cite{fagnp}.  Since then, a lot of work has been
achieved
about the logical definability of classes of formal languages of finite or
infinite words,
or of relational structures like graphs,
see \cite{fag}  \cite{pin}  \cite{tho96} \cite{imm}
for a survey about this field of research.

 Several extensions of existential monadic second order logic over words
have been studied.

 Lauteman,  Schwentick and  Therien proved  that context free languages are
characterized by existential  second order formulas in which  the second
order quantifiers
bear only on  matchings, i.e. pairing relations without crossover,
\cite{lst}.

  Parigot and Pelz, and more recently Yamasaki,
 extended monadic second order logic with two second order relation symbols
and characterized classes of Petri net ($\om$)-languages \cite{pp85}
\cite{pel} \cite{yam}.

  Eiter,  Gottlob and Gurevich  studied the relationship between monadic
second order logic
and syntactic fragments of existential second order logic over (finite)
words \cite{egg}.
Distinguishing  prefix classes, they determined which of them define only
regular languages and
which of them have the same expressive power as monadic second order logic.

  Another extension has been introduced by Ressayre, in order to apply some
powerful tools
of model theory to the  study of formal languages, \cite{ress}.
He defined  locally finite sentences (firstly called local). A  locally
finite sentence $\vp$
is a  first order  sentence which is equivalent to
a universal one and whose models satisfy simple structural properties:
closure under functions
takes a finite number $n_\vp$ of steps.
\nl These  syntaxic and semantic restrictions  allow a meaningful  use of
the  notion
of indiscernables and lead to beautiful stretching theorems connecting  the
existence
of some well ordered infinite models of $\vp$ with the existence of some
finite
models generated by indiscernables \cite{fr}.

 Locally finite languages are defined by second order formulas in the form
$\exists \bar{R} ~~\exists \bar{f} ~~\vp$ where $\vp$ is a locally finite
sentence and $\bar{R}$
(respectively, $ \bar{f}$) represent the relation  (respectively, function)
symbols
in the signature of $\vp$.
\nl  These second order quantifications are much more
general than the monadic ones as the following results show:
\begin{itemize}
\ite  Each regular language is locally finite, \cite{ress},
 and many context free as well as
non context free \la s are locally finite  \cite{loc}.

\ite  Each \orl~  is a
locally finite  $\om$-language, \cite{loc} \cite{of},
 and there exist many more locally finite \ol s as we shall see below.

\ite  This result  is extended to languages of transfinite length words:  if
$\alpha$ is
 an ordinal $<\om^\om$,  each regular
  $\alpha$-language
is also  locally finite  \cite{loc}.
\end{itemize}

\noi  But a pumping lemma, following from a  stretching theorem,  makes
locally finite \ol s  keep important properties of \orl s, \cite{ress}
\cite{fr}.
It is an analogue for each locally finite
\ol~ of the property:
\begin{center}
`` A \orl~ is non empty if and only if it contains an ultimately periodic
word ".
\end{center}
This lemma implies in a similar manner the decidability of the emptiness
problem for
locally finite \ol s.  Moreover for each countable ordinal $\alpha
<\om^\om$, the
decidability of  the emptiness problem for locally finite $\alpha$-languages
follows from
similar arguments,  \cite{fr}.
\nl Other decidability results, as the decidability of the problem: ``is a
given
finitary locally finite language infinite?"
follow from stretching theorems of \cite{ress}\cite{fr}.

  These interesting properties of locally finite languages naturally lead to
the question
of the richness of the class of locally finite languages: how large is this
class? What are
its closure properties?

 The study of locally finite languages of finite words was begun by Ressayre
in \cite{ress}
and continued in  \cite{loc}.
 We focus  in this paper on the class $LOC_\om$ of
 locally  finite \ol s and study  classical closure
properties for this class.  In particular, we show that $LOC_\om$
is  neither  closed under intersection, nor under complementation.
The proof uses the notion of rational cone of finitary languages which is
important in formal language theory  and
the notion of indiscernables in a structure, often used in model theory.
\nl This gives an answer to a question of Ressayre, \cite{ress2}.
Of course we would have preferred a positive answer to this question which
would have provided
a useful class of sentences for specification and verification of properties
of non-terminating
systems.  But this leaves still open, for further study,
the possibility to find such a useful class of sentences as
a subclass of the class of locally finite sentences.

 In section 2, we give the first definitions and
some examples of locally finite \ol s.
In section 3,  closure properties for \ol s are investigated.
We show that the class $LOC_\om$ is not closed under intersection with \orl
s thus $LOC_\om$
is  neither  closed under intersection, nor under complementation (because
$LOC_\om$ is closed
under union).
Then we prove that $LOC_\om$ is closed under $\lambda$-free morphism and
$\lambda$-free substitution of locally finite (finitary) languages.

\section{Definitions and examples}
\subsection{Definitions}
We briefly indicate now some basic facts about first order logic and model
theory.
See for example \cite{ck} for more background on this subject.

  We consider here formulas of first order logic. The language of first
order logic contains
(first order) variables x, y, z, $\ldots$  ranging over
elements of a structure, logical symbols:
 the connectives
$\wedge$~(and),  $\vee$~ (or),  $\ra$~ (implication), $\neg$~ (negation),
and the
 quantifiers $\fa$~ (for all), and $\exists$~
(there exists), and also the binary predicate symbol of identity $=$.
\nl A signature is a set of constant, relation ( different from = ) and
function symbols.
 we shall consider here only finite signatures.

 Let $Sig$ be a finite signature. We define firstly the set of terms in the
signature $Sig$
 which is built inductively as follows:
\begin{enumerate}
\ite A variable is a term.
\ite A constant symbol is a term.
\ite If $F$ is a m-ary function symbol and $t_1, t_2, \ldots , t_m$ are
terms, then
$F(t_1,  \ldots , t_m)$ is  a term.
\end{enumerate}

\noi We then define the set of atomic formulas which are in the form given
below:
\begin{enumerate}
\ite If $t_1$ and $t_2$ are terms, then $t_1=t_2$ is an atomic formula.
\ite If $t_1, t_2,  \ldots , t_m$~ are terms and $R$ is a m-ary relation
symbol, then
$R(t_1, \ldots , t_m)$ is  an atomic formula.
\end{enumerate}

\noi Finally the set of formulas is built  inductively from atomic formulas
as follows:
\begin{enumerate}
\ite An atomic formula is a formula.
\ite If $\vp$ and $\psi$ are formulas, then $\vp \wedge \psi$, $\vp \vee
\psi$,
$\vp \ra \psi$ and $\neg \vp$ are formulas.
\ite If $x$ is a variable and $\vp$ is a formula, then $\fa x \vp$ and
$\exists x \vp$ are
formulas.
\end{enumerate}

\noi An open formula is a formula with no  quantifier.
\nl  We assume the reader to know the notion of free and bound occurrences
of a variable in a
formula. Then a sentence is a formula with no free variable.
\nl A sentence  in prenex normal form is in the form
$\vp = Q_1x_1  \ldots Q_nx_n \vp_0(x_1, \ldots , x_n)$,
where each $Q_i$ is either the quantifier $\fa$ or the quantifier $\exists$
and the formula
$\vp_0$ is an open formula.
\nl It is well known that every sentence is equivalent to a sentence written
 in prenex normal form.
\nl A sentence is said to be universal if it is in prenex normal form and
each quantifier
is the universal quantifier $\fa$.

 We then recall the notion of a structure in a signature $Sig$:
A  structure is in the form:
 $$M = (|M|, (a^M)_{a \in Sig} ) $$
\noi Where $|M|$ is a set called the universe of the structure, and for
$a\in Sig$,
 $a^M$ is the interpretation of $a$ in $M$:
\nl If $f$ is a m-ary function symbol in $Sig$, then $f^M$ is a function:
$M^m \ra M$.
\nl If $R$ is a m-ary relation  symbol in $Sig$, then $R^M$ is a relation:
$R^M \subseteq M^m$.
\nl If $a$ is a constant symbol in $Sig$, then $a^M$ is a distinguished
element in $M$.

 In order  to simplify the notations we shall sometimes write $a$ instead of
$a^M$ when the
meaning is clear from the context.

 When $M$ is a structure and $\vp$ is a sentence in the same signature
$Sig$, we write
$M \models \vp$ for `` $M$ is a model of $\vp$ ", which means that $\vp$ is
satisfied in the structure
$M$. A detailed exposition of  these notions may be found in  \cite{ck}.

 When $M$ is a structure in the signature $Sig$ and $Sig_1$ is another
signature such that
$Sig_1 \subseteq Sig$, then the reduction of $M$ to the signature $Sig_1$ is
denoted $M|Sig_1$.
It is a structure in the signature $Sig_1$ which has same universe $|M|$ as
$M$, and the same
 interpretations for symbols
in $Sig_1$.
Conversely an expansion of a structure $M$ in the signature $Sig_1$ to a
structure in the signature
$Sig$ has same universe as $M$ and same interpretations for symbols
in $Sig_1$.

 When $M$ is a structure in a signature $Sig$ and $X\subseteq |M|$, we
define:
\nl $cl^1(X, M)=X \cup \bigcup_{\{f ~{\rm n-ary~ function~ of ~} Sig~\} }
~f^M(X^n)
\cup \bigcup_{\{a ~{\rm ~ constant~ of ~} Sig~\} } a^M $
\nl $cl^{n+1}(X, M)=cl^1(cl^n(X, M), M) \quad {\rm ~ for ~an ~integer~}
n\geq 1$
\nl  and $cl(X, M)=\bigcup_{n\geq 1} cl^n(X, M)$ is  the closure of $X$ in
$M$.

 Let us now define locally finite sentences. We shall denote S($\varphi$)
the
signature of a first order sentence  $\varphi$, i.e. the set of
 non logical symbols  appearing in  $\varphi$.

\begin{Deff}\label{defloc} A first order sentence  $\varphi$
 is locally finite if and only if:
\begin{enumerate}
\item[a)] $M\models\varphi$ and  $X\subseteq|M|$ imply
$cl(X,M)\models\varphi$
\item[b)] $\exists n\in \mathbb{N}$ such that  $\forall M$,
if $ M\models \varphi$ and  $X \subseteq |M|$,
then $ cl(X,M)=cl^n(X, M)$  (closure in models of  $\varphi$ takes less than
$n$ steps).
\end{enumerate}
\end{Deff}

\noi {\bf Notation.} For a locally finite sentence $\varphi$, we shall
denote by
 $n_\varphi$  the
 smallest integer $n\geq 1$ satisfying  b) of the above definition.

\begin{Rem} Because of a) of Definition \ref{defloc}, a locally finite
sentence
 $\varphi$ is always equivalent to a universal
sentence.
\end{Rem}

\noi We  now introduce some notations for finite or infinite words.
\nl Let $\Si$ be a finite alphabet whose elements are called letters.
A finite word over $\Si$ is a finite sequence of letters:
 $x=a_0 \ldots a_n$ where $\fa i\in [0; n]$ $a_i \in\Si$.
 We shall denote $x(i)=a_i$ the $i+1^{th}$ letter of $x$
and $x[i]=x(0)\ldots x(i)$ for $i\leq n$. The length of $x$ is $|x|=n+1$.
The empty word will be denoted by $\lambda$ and has 0 letter. Its length is
0.
 The set of finite words over $\Si$ is denoted $\Sis$.
 $\Si^+ = \Sis - \{\lambda\}$ is the set of non-empty words over $\Si$.
 A (finitary) language $L$ over $\Si$ is a subset of $\Sis$.
 Its complement (in $\Sis$) is $L^-= \Sis - L$.
 The usual concatenation product of $u$ and $v$ will be denoted by $u.v$ or
just  $uv$.
 The set of non negative integers is denoted by $\mathbb{N}$.
 For $V\subseteq \Sis$, we denote  \quad
$ V^\star = \{ v_1\ldots v_n  \mid  \fa i \in [1; n]
\quad v_i \in V \quad \} \cup \{\lambda\}$.

 The first infinite ordinal is $\om$.
\nl An $\om$-word over $\Si$ is an $\om$-sequence $a_0a_1 \ldots a_n \ldots
$, where $\fa i\geq 0$~
$a_i \in\Sigma$.
\nl For  $\sigma \in \Sio$, $\sigma(n)$ is the $n+1^{th}$ letter of $\sigma$
and
 $\sigma[n]=\sigma(0)\sigma(1)\ldots \sigma(n)$.
\nl The set of $\om$-words over  the alphabet $\Si$ is denoted by $\Si^\om$.
 An  $\om$-language over $\Sigma$ is a subset of  $\Si^\om$.
The $\om$-power of a finitary language $V\subseteq \Sis$  is the
$\om$-language
  $V^\om = \{ \sigma =u_1\ldots u_n\ldots  \in \Si^\om \mid  \fa i\geq 1 ~~
u_i\in V \}$.
\nl  For a subset $A\subseteq \Si^\om$, the complement of $A$ (in $\Si^\om$)
is
$\Si^\om - A$ denoted $A^-$.
\nl The concatenation product is extended to the product of a
finite word $u$ and an $\om$-word $v$: the infinite word $u.v$ is then the
$\om$-word
satisfying:
\nl $(u.v)(k)=u(k)$  if $k < |u|$ , and  $(u.v)(k)=v(k-|u|)$  if $k \geq
|u|$.

 A word over $\Si$ may be considered as a structure in the  usual manner:

  Let $\Sigma$ be a finite  alphabet. For each letter
 $a \in \Sigma$ we denote  $P_a$ a unary predicate
and   $\Lambda _\Sigma$ the  signature $\{ <, (P_a)_{a\in \Sigma} \}$.
The  length $|\sigma |$ of a non-empty finite word
$\sigma \in  \Sigma^\star$ may be written
$|\sigma |=\{0, 1,\ldots , |\sigma |-1 \} $. $\sigma$ is
 identified to the structure $( |\sigma|, <^\sigma , (P_a^\sigma)_{a \in
\Sigma} )$
 of  signature
 $\Lambda _\Sigma $ where
\nl  $P_a^\sigma = \{ i<|\sigma| \mid \mbox{  the  } i+1^{th} \mbox{  letter
of }
  \sigma \mbox{  is an  } a \}$.

 In a similar manner if $\sigma$ is  an $\om$-word over the alphabet
$\Sigma$,
then $\om$ is the length of the word $\sigma$ and we may write
$|\sigma |= \om =\{0, 1, 2, 3, \ldots \}$.
$\sigma$ is
 identified to the structure $( \om , <^\sigma , (P_a^\sigma)_{a \in
\Sigma} )$
 of  signature
 $\Lambda _\Sigma $ where
\nl  $P_a^\sigma = \{ i<\om \mid \mbox{  the  } i+1^{th} \mbox{  letter of }
  \sigma \mbox{  is an  } a \}$.

\begin{Deff} Let  $\Sigma$ be a finite alphabet  and  $L\subseteq
\Sigma^\star$.
\nl Then [$L$ is a locally finite language ] $\longleftrightarrow$

 there exists a locally finite sentence
 $\varphi$ in a signature $\Lambda \supseteq \Lambda_\Sigma$ such
 that  $\sigma \in L$ iff  $ \exists M, M\models \varphi $ and $ M|\Lambda
_\Sigma =\sigma$].
\nl We then denote  $L=L^\Sigma(\varphi)$  and, if there is no ambiguity,
 $L=L(\varphi)$ the locally
finite  language defined by $\varphi$.
\nl The class of locally finite languages will be denoted $LOC$.
\end{Deff}

\noi
The empty word  $\lambda$  has 0 letters. It is represented by the empty
structure.
Recall that if  $L(\varphi)$ is a locally finite language then
$L(\varphi)-\{\lambda\}$ and
$L(\varphi)\cup \{\lambda \}$ are also   locally finite  \cite{loc}.

\begin{Deff} Let  $\Sigma$ be a finite alphabet  and  $L\subseteq
\Sigma^\om$.
\nl Then  [ $L$ is a locally finite $\om$-language ] $\longleftrightarrow$
[ there exists a locally finite sentence
 $\varphi$ in a signature $\Lambda \supseteq \Lambda_\Sigma$ such
 that $\fa \sigma \in \Sio$~ $\sigma \in L$ iff
$ \exists M, M\models \varphi $ and $ M|\Lambda _\Sigma =\sigma$ ].
\nl We then denote  $L=L_\om^\Sigma(\varphi)$, and, if there is no
ambiguity,
 $L=L_\om(\varphi)$ the locally
finite  $\om$-language defined by $\varphi$.
\nl The class of locally finite $\om$-languages will be denoted $LOC_\om$.
\end{Deff}

\begin{Rem}
 The notion of  locally finite ($\om$)-language  is very different of the
usual
 notion of local ($\om$)-language which represents a subclass of the class
of rational
($\om$)-languages.
 But from now on, as in \cite{loc}, things being well defined and precised,
we shall  call simply local ($\om$)-languages (respectively, local
sentences)
the locally finite ($\om$)-languages (respectively, locally finite
sentences).
\end{Rem}

\subsection{Examples of local \ol s}

The following example should not be skipped because it is crucial to Theorem
\ref{theokcloc}
below.

\Exab \label{wordloc}
The \ol~ which contains only the word $\sigma =  abab^2ab^3ab^4\ldots $
(where the $i$-th occurrence of $a$ is followed by the factor $b^{i}a$)
is a local \ol~ over the alphabet $\{ a, b\}$.
\Exae

\proo Let the signature $S(\vp)=\{ P_a, P_b, <, p, p', f\}$, where $p$ and
$p'$ are unary
function symbols, $f$ is a binary function symbol. And let $\vp$ be the
following sentence,
conjunction of:

\begin{enumerate}
\ite[(1)]  $\fa xyz [ (x\leq y \vee y\leq x) \wedge (x\leq y \wedge  y\leq x
\lra x=y) \wedge
(x\leq y \wedge y\leq z \ra x\leq z) ]$,
\ite[(2)]   $\fa x [ P_a(x) \lra \neg P_b(x) ]$,
\ite[(3)]   $\fa xy [ ( x<y \wedge P_a(x) \wedge P_a(y) ) \ra
 P_b(f(xy)) ]$,
\ite[(4)]   $\fa xy [ x\geq y \ra f(xy)=x ]$,
\ite[(5)]   $\fa xy [ \neg P_a(x) \vee \neg P_a(y) \ra f(xy)=x ]$,
\ite[(6)]   $\fa x [ P_a(p(x)) \wedge P_a(p'(x)) ]$,
\ite[(7)]  $\fa x [ \neg P_a(x) \ra p(x)<p'(x) ]$,
\ite[(8)]   $\fa x [ P_a(x) \ra p(x)=p'(x)=x ]$,

\ite[(9)]   $\fa x [ \neg P_a(x) \ra x=f(p(x)p'(x)) ]$,
\ite[(10)]   $\fa xx'y \in P_a [x<x'<y \ra f(x'y)<f(xy)<y ]$,
\ite[(11)]   $\fa xyy' \in P_a [x\leq y'<y \ra y'<f(xy)<y ]$.

\end{enumerate}

\noi Above sentence $(1)$  means that   ``$<$ is a linear order ''. The
sentence $(2)$ expresses
that $P_a, P_b$ form a partition in any  model $M$ of $\vp$. The sentence
$(3)$
says that $f$ is a function from
$\{(x, y) \mid  x<y \wedge P_a(x) \wedge P_a(y) \}$ into $P_b$
while sentences $(4)$-$(5)$ express
that the function $f$ is trivially defined elsewhere.
\nl Sentences $(6)$-$(7)$ say that $p$ and $p'$ are functions defined from
$\neg P_a=P_b$
into $P_a$ and $(8)$ states that $p$ and $p'$ are trivially defined on
$P_a$.
\nl The projections $p$ and $p'$ are used to  say that the
function $f$ is surjective from $\{(x, y) \mid  x<y \wedge P_a(x) \wedge
P_a(y) \}$
onto $P_b$; this is implied by sentence $(9)$.
\nl
The $10^{th}$ and $11^{th}$  conjunctions are used to order the elements
of $f(P_a\times P_a)$ in order
to obtain the word $\sigma =  abab^2ab^3ab^4\ldots $ when the reduction to
the signature of
words is considered.
\nl Notice that $(10)$-$(11)$ imply also that $f$ is injective hence is in
fact a
bijection from
$\{(x, y) \mid  x<y \wedge P_a(x) \wedge P_a(y) \}$ into $P_b$.
\ep

\begin{Rem}
We have defined the functions $f$ and $p$, $p'$, in a trivial manner (like
$f(xy)=x$ or $p(x)=x$) where they were not useful for defining the local
\ol~ $\{\sigma\}$,
(see the  conjunctions $(4)$, $(5)$ and $(8)$).
This will imply that closure in models of $\vp$ takes at most a finite
number of steps.
This method will be applied in the construction of other local sentences in
the sequel
of this paper.
\end{Rem}

\noi We can easily check that $\vp$ is equivalent to a universal formula
and that closure in its models takes
at most $n_\vp = 2$ steps:  one takes closure under the functions $p, p'$
then by $f$.
\nl Hence $\vp$ is a local sentence and by construction:
  $L_\om^{\{a, b\}}(\vp)=\{abab^2ab^3\ldots \}$.

Let us give some examples of closure in a model $M$ of $\vp$ such that
$M|\Lambda_{\{a, b\}}=\sigma=abab^2ab^3ab^4\ldots $

 Let $X_n \subseteq |M|$ be the segment of $M$ corresponding to the
segment $ab^na$ of $\sigma$. Then the closure of $X_n$ under the functions
$p, p'$ is the set
$X_n \cup Z_n$ where $Z_n$ corresponds to the set of the $(n-1)$ first
letters $a$ of $\sigma$.
$cl(X_n, M)$ is the closure of $X_n \cup Z_n$ under $f$  and it is
the initial segment of $M$ corresponding to the
initial segment $abab^2ab^3ab^4\ldots ab^na$ of $\sigma$.

Let now $Y \subseteq |M|$ be the segment of $M$ corresponding to the three
last
letters $b^2a$ of the segment $ab^na$  of $\sigma$, for some integer $n\geq
3$.
Then the closure of $Y$
under the functions $p, p'$ is $Y \cup Z$ where $Z$ corresponds to the set
of the
two first letters $a$ of $\sigma$.
The closure of $Y \cup Z$  under $f$ is the  set
$cl(Y,M)$ which induces the word $abab^2a$ but which is not a segment of $M$
because
it contains  the two first letters $a$ and the $(n+1)$-th letter $a$ of
$\sigma$ but not
any other letter $a$ of $\sigma$.

 We are going now to get more examples of local \ol s. Recall first the
following:

\begin{Deff}
The $\om$-Kleene closure of a family $\mathcal{L}$  of  finitary \la s
is:
$$\om\mbox{-KC}(\mathcal{L}) = \{ \cup_{i=1}^n U_i.V_i^\om  \mid  \fa i\in
[1, n] ~~
 U_i, V_i \in \mathcal{L} \}$$
\end{Deff}

\noi This notion of $\om$-Kleene closure appears in the characterization of
the class
$REG_\om$
of  \orl s (respectively,  of the class $CF_\om$ of context free \ol s)
 which is  the $\om$-Kleene closure
of the family $REG$ of regular finitary languages (respectively,   of the
family $CF$ of context free finitary languages), \cite{tho} \cite{pp}
\cite{sta}.

\noi A natural question arises: does a similar characterization hold for
local \la s?
The answer is given by the following:

\begin{The}\label{theokcloc}
The $\om$-Kleene closure of the class $LOC$ of finitary local \la s is
strictly included
into the class $LOC_\om$ of local \ol s.
\end{The}

\proo
We have already proved that $\om\mbox{-}KC(LOC) \subseteq LOC_\om$ in
\cite{loc}.
In order to show that the inclusion is strict, remark that if an
$\om$-language $L$ belongs to
$\om\mbox{-}KC(LOC)$, then $L$ contains at least an ultimately periodic
word,
i.e. a word in the form
$u.v^\om$ where $u$ and $v$ are finite words.
Now we can easily check that the local \ol~ given in example \ref{wordloc}
does not contain any ultimately periodic word because its single word is not
ultimately periodic.  \ep

 A first consequence of Theoreme \ref{theokcloc} is that
every regular \ol~ is a local \ol, i.e.  $REG_\om \subseteq LOC_\om$,
because
every finitary regular language is local \cite{ress}.
\nl We had shown in \cite{loc} that many context free languages are local
thus  $CF_\om = \om\mbox{-}KC(CF)$ implies that many context free \ol s are
local.
The problem to know whether each context free language is local is still
open
but by Theorem \ref{theokcloc}, $CF \subseteq LOC$ would imply that
$CF_\om \subseteq LOC_\om$.

  The $\om$-language given in example \ref{wordloc} is local but non context
free because
every context free $\om$-language contains at least one ultimately periodic
word.
\nl We proved in \cite{loc} that the finitary language
$U=\{a^nb^{n^2}  \mid  n\geq 1 \}\subseteq \{a, b \}^\star$
is  local. Thus the $\om$-language $U.c^\om \subseteq \{a, b, c\}^\om$ is
local
but  $U.c^\om$ is  not  context free  because $U \notin CF$, \cite{cg}.
\nl These two examples show that the inclusion $LOC_\om  \subseteq  CF_\om$
does not hold.

\section{Closure properties of locally finite omega languages}

\begin{The}\label{theclo}  The class of  locally finite omega languages  is
not closed under
  intersection
 with a regular \ol~ in the form
  $L.a^\omega$ where  $L$ is a rational language,
$L\subseteq \Sigma^\star$ and  $a\notin \Sigma$.
Hence $LOC_\om$ is neither closed under  intersection nor under
complementation.
\end{The}

\noi To prove this theorem, we shall proceed by successive lemmas.
We shall assume that every language
considered here is constituted of words over a finite alphabet included in
a given countable set $\Sigma_D$.

 We firstly define the family  $I$ of finitary languages by:  for a finitary
language
  $L\subseteq \Sigma^\star$,  where
 $\Sigma \subseteq \Sigma_D$ is a finite alphabet,   $L\in I$ if and only if
 $L.a^\omega \in LOC_\omega$ whenever $a$ is a letter of
  $\Sigma_D-\Sigma$.

It is easy to see that if  $L.a^\omega$ is a local  \ol~ and
 $a\in \Sigma_D-\Sigma$, then for all $b \in \Sigma_D-\Sigma$, it holds that
 $L.b^\omega \in LOC_\omega$. It suffices to replace the predicate $P_a$ by
$P_b$ in the
sentence defining  $L.a^\omega$.

\begin{Lem}\label{clolem1} $I$ is closed under inverse alphabetic morphism.
\end{Lem}

\proo  Let  $L\in I$ , i.e.  $L\subseteq \Gamma^\star$  for some finite
alphabet $\Ga$,
 $a \notin \Gamma$ and
 $L.a^\omega=L_\omega^{\Gamma\cup \{a\} }(\varphi)$ for a local sentence
 $\varphi$.
\nl Let $h$ be the alphabetic morphism: $\Sigma^\star \rightarrow
\Gamma^\star$ , defined by
 $h(c)\in \Gamma\cup \{\lambda \}$  for $ c \in \Sigma$,  where $\lambda$ is
the empty word.
And let  $\Sigma'= \{ c\in\Sigma \mid h(c)=\lambda\} $.
We assume, possibly changing $a$, that
 $a\notin \Sigma$.
\nl We first replace in  $\varphi$ the letter predicates  $(P_c)_{c\in
\Gamma}$ by
 $(Q_c)_{c\in \Gamma} $.

 The language  $h^{-1}(L).a^\omega$ is then defined by the following
sentence $\psi$,
in the   signature
$S(\psi)=\{ P, A, (P_c)_{c\in \Sigma }, P_a\} \cup S(\varphi )$,
where $S(\varphi )$ contains the letter predicates  $Q_c$ for  $c
\in\Gamma\cup\{a\}$, $P$
 is a unary predicate symbol and   $A$ is a constant symbol.
$\psi$ is the conjunction of:

\begin{itemize}

\ite ( $ < $ is a linear order ),

\ite $ ( (P_c)_{c\in \Sigma}, P_a )$ form a  partition,

\ite  $\forall x_1...x_n \in P [\varphi_0(x_1...x_n) \wedge\bigwedge
_{c\in\Gamma} (Q_c(x_1)
\leftrightarrow \bigvee_{d\in h^{-1}(c) } P_d(x_1 ) ) \wedge (Q_a(x_1)
\leftrightarrow
 P_a(x_1) ) ]$, where  $\varphi =\forall x_1...x_n \varphi_0(x_1...x_n)$
with
 $\varphi_0$ an open formula,

\ite $\forall x_1...x_k [ ( \bigvee _ { 1\leq j \leq k } \neg P(x_j) )
\rightarrow
 g(x_1...x_k)=x_1 ]$ , for each k-ary function $g$ of  $S(\varphi )$,

 \ite $(P_c)_{c\in \Sigma '}$ form a partition of  $\neg P$,

\ite $P(B)$, for each constant  $B$ of  $S(\varphi )$,

\ite $\forall x y [\neg P_a(y) \wedge P_a(x) \rightarrow y < x ]$,

\ite $P_a(A)$.

\end{itemize}

\noi  $\psi$ is  equivalent to a universal sentence   and closure in its
models takes at most
 $n_\varphi +1$ steps. Hence  $\psi$ is  local and by construction
$L_\omega^ {\Sigma\cup\{a\} }(\psi) =h^{-1}(L).a^\omega$ .\ep

\begin{Lem}\label{clolem2} $I$ is closed under non erasing alphabetic
morphism.
\end{Lem}

\noi Recall that an alphabetic morphism  $h:
\Sigma^\star\rightarrow\Gamma^\star$,
defined by  $h(c)\in\Gamma\cup\{\lambda\}$, for $c\in\Sigma$ is said to be
non  erasing if
$\forall c \in \Sigma,  h(c)\in\Gamma$.

\proo  Let $L\in I, L\subseteq\Sigma^\star$.
Let  $a\notin \Sigma$ and  $L.a^\omega=L_\omega^{\Sigma\cup\{a\} }(\varphi)$
for a local sentence  $\varphi$.
Let $h$ be a  non erasing alphabetic morphism given by $h:
\Sigma\rightarrow\Gamma$.
\nl Moreover we  assume that  $a\notin \Gamma $ (possibly changing $a$).
Then the
language  $h(L).a^\omega$
is defined by the following formula  $\psi$, in the signature
 $S(\psi)=S(\varphi)\cup \{(Q_c)_{c\in\Gamma} \}$. The sentence $\psi$ is
the
conjunction of:

\begin{itemize}

\ite  $\varphi$,

\ite   $\forall x [\bigwedge_{c\in\Sigma} (P_c(x) \rightarrow
Q_{h(c)}(x) ) ]$,

\ite   $[ (Q_c)_{c\in\Gamma}, (P_a) ]$  form a  partition.

\end{itemize}

\noi
\nl $\psi$ is  local and if the predicates  $(Q_c)_{c\in\Gamma}, P_a$, are
the letter
predicates,  $\psi$ defines the \ol~
$L_\omega^ {\Ga\cup\{a\} }(\psi) =h(L).a^\omega$ .  \ep

\begin{Lem}\label{clolem3} $I$ contains the finitary local languages.
\end{Lem}

\proo
$LOC_\om$ contains the omega  Kleene closure of the class $LOC$ of
finitary local languages, and for each letter $a$ the language $\{a\}$ is
local.  \ep

 Recall now the definitions of the Antidyck language and of a rational cone
of languages.

\begin{Deff} The  Antidyck language over  two sorts of parentheses is the
language
$Q_2'^\star=\{ v\in (Y\cup \bar{Y})^\star \mid  v\rightarrow^\star \lambda
\}$,
where  $Y=\{y_1, y_2\}$, $\bar{Y}=\{\bar{y_1}, \bar{y_2}\}$  and
$\rightarrow^\star$ is
 the transitive closure of
 $\rightarrow$ defined in  $(Y\cup \bar{Y})^\star $ by:
\nl $\forall y\in Y ~~~~  yv_1\bar{y}v_2\rightarrow v_1v_2$
~~if and only if ~~ $v_1\in Y^\star$.
\end{Deff}

\noi
The Antidyck language $Q_2'^\star$ may be seen as the language containing
 words with two sorts of parentheses, such that:
``the first  parenthesis to be opened  is the first to be closed"

\begin{Deff}[\cite{bers}]
A rational cone is a class of \la s which is closed under morphism, inverse
morphism, and
intersection with a rational language. (Or, equivalently to these three
properties, closed
under rational transduction).
\end{Deff}

\noi The notion of rational cone has been much studied.
In particular the Antidyck language $Q_2^{'\star}$ is a generator of the
rational
cone of the recursively enumerable languages, \cite{fzv}.
\nl On the other hand Nivat's Theorem states that
a class of languages which is closed under alphabetic morphism, inverse
alphabetic  morphism, and intersection with a rational language, is a
rational cone,
\cite{bers}.  Moreover every rational  transduction $t$
is in the form $t(u)=g[h^{-1}(u)\cap R]$, where $ g$ et $h$ are alphabetic
morphisms
and $R$ is a rational language. Thus
every recursively enumerable language may be written in the form
$g[h^{-1}(Q_2'^\star)\cap R ]$, where  $R$ is a rational language, $g$ and
$h$
are alphabetic morphisms.
\nl This result will be used here because the  language $Q_2'^\star$ is
local, \cite{loc}.

 Return now to the proof of Theorem \ref{theclo} and suppose that
 $LOC_\om$ were closed under intersection with the
  languages $R.a^\omega$, where $R\subseteq \Sigma^\star$ is a
 rational language and  $a\notin \Sigma$.

\begin{Cla}\label{cl1}
$I$ would be  closed under intersection with a  rational language.
\end{Cla}

\proo Let  $L\in I$,  $L\subseteq \Sigma^\star$,
 $R \subseteq \Sigma^\star$ be a  rational language
and  $a\notin \Sigma$.  $L.a^\omega$ is  a local \ol~
 because  $L\in I$ and
 $(L.a^\omega)\cap(R.a^\omega)=(L\cap R).a^\omega$ would be  a local
$\omega$-language,
 hence by definition of $I$, $L\cap R$ would belong to  $I$.  \ep

\begin{Cla}\label{cl2} $I$ would contain
every language  in the form
$g[h^{-1}(Q_2'^\star)\cap R ]$, where  $h$ is an alphabetic morphism, $g$
 is a non erasing alphabetic morphism and  $R$ is a rational language.
\end{Cla}

\proo  It follows from the  lemmas \ref{clolem1}, \ref{clolem2},
\ref{clolem3}, the fact that
$Q_2'^\star$ is local and Claim \ref{cl1}.  \ep

\begin{Cla}\label{cl3}    There would exist an
 erasing alphabetic morphism  $h$  and $L\in I$ such that
$\{0^n1^p \mid  p>2^n\}=h(L)$, where an alphabetic morphism is said
 to be erasing if it is in the form
$h: \Sigma \rightarrow \Sigma\cup \{\lambda\}$, with  $h(c)=c$ if  $c\in A$
and  $h(c)=\lambda$ if
 $c\in \Sigma -A$, for some subset $A\subseteq \Sigma$.
\end{Cla}

 \proo    We know that every recursively enumerable language may be written
in the form
$g[h^{-1}(Q_2'^\star)\cap R ]$, where  $R$ is a rational language, $g$ and
$h$
are alphabetic morphisms.
\nl But every  alphabetic morphism
  may be  obtained
as composed firstly  by a non erasing alphabetic morphism followed by an
erasing alphabetic morphism.
\nl Thus it follows from  Claim \ref{cl2} that
  each  recursively enumerable language, and in particular the  language
$\{0^n1^p \mid  p>2^n\}$, would be
 the image by an erasing alphabetic morphism
 of a language of  $I$.
 \ep

Let then  $h$ be an erasing morphism
 $\Sigma^\star \rightarrow \Sigma^\star$  where $\{0, 1\}\subseteq \Sigma,
h(0)=0$ and
$h(1)=1$ and  $h(c)=\lambda$ if  $c\in \Sigma -\{0,1\}$ and let
 $L\subseteq \Sigma^\star$ be a language
  such that  $h(L)= \{0^n1^p \mid  p>2^n\}$.
\nl Assume that $L$ belongs to $I$, so if $a\notin \Sigma$,  $L.a^\omega$ is
 a local \ol~ and   there exists a local sentence  $\varphi$ such that
$L.a^\omega=L_\omega^{\Sigma\cup\{a\} }(\varphi)$.
\nl For all $n\geq 1$ let   $M_n$ be a model of  $\varphi$ of  order type
$\omega$
such that  $M_n\mid\Lambda_{\Sigma\cup\{a\} }=\sigma_n.a^\omega$,  where
$\sigma_n\in\L$
and the number of occurrences  of  $0$ in $\sigma_n$ is $n$ and the number
of occurrences  of
 $1$ in
 $\sigma_n$  is  $p_n >2^n$.
\nl Let us now set the following definition in view of next lemma.

\begin{Deff} Let  $X$ be a set included in a  structure $M$ and
$P\subseteq |M|$. $X$ is a set of indiscernables above  $P$ for the
atomic formulas   of complexity  $\leq k$, i.e.
 whose terms result by at most
$k$ applications of function symbols, for $k\in \mathbb{N}$,  if and only
if:
\begin{enumerate}
\item[i)] $X$ is linearly ordered by  $<$.
\item[ii)]  Whenever  $\bar x$ and  $\bar y$ are some n-tuples of elements
of  $X$ which
are isomorphic for the order of
 $(X, <)$, $\bar x$ and  $\bar y$ satisfy the same atomic formulas   of
complexity  $\leq k$,
 with parameters in  $P$.
\end{enumerate}
\end{Deff}

\begin{Lem}  In  the above conditions where $M_n$ is defined for every
integer $n \geq 1$,
 there exists in $M_n$ an infinite set  $X_n$ of indiscernables above
$P_0^{M_n}$
for the
atomic formulas   of complexity    $\leq n_\varphi$, with  $X_n\subseteq
P_a^{M_n}$
\end{Lem}

\proo  Let  $m(\varphi)$ be the maximum number of variables of
the
atomic formulas   of complexity
 $\leq n_\varphi$ i.e.  whose terms result by at most
$n_\varphi$  applications of function symbols. These terms form a finite set
 $T_\varphi$ .

 For all strictly increasing sequences  $\bar x$ and  $\bar y$ of length
$m(\varphi)$  of
$P_a^{M_n}$ , let us set  $\bar x \sim \bar y$  if and only if $\bar x$ and
$\bar y$
satisfy in
 $M_n$
the same atomic formulas  with parameters in   $P_0^{M_n}$ and of
complexity  $\leq n_\varphi$.

   $P_0^{M_n}$ is a finite set of  cardinal $n$, hence
the set of atomic formulas  with parameters in
 $P_0^{M_n}$ and of complexity $\leq n_\varphi$ is also finite.

 Then applying the infinite Ramsey Theorem, we can find $X_n\subseteq
P_a^{M_n}$
 homogeneous for  $\sim$ and  infinite. This is the set we are looking for.
\ep

 We return now to the proof of Theorem \ref{theclo} and
 consider in  $ |M_n| $ the subset $X_n \cup P_0^{M_n} =Y_n$.
 This subset is infinite  hence it is of order type  $\omega$ in $M_n$ and
it
generates in $M_n$
a model of order type   $\omega$ too, which will be denoted by
$M_n(Y_n)=A_n$.

 This model of  $\varphi$ induces a word $u_n.a^\omega$ of  $L.a^\omega$,
such that
there are  in  $u_n$:
 $n$ occurrences of the letter $0$ and  $q_n\leq p_n$  occurrences of the
letter $1$.
 But  $u_n.a^\omega \in L.a^\omega$ implies that $2^n< q_n$ also holds.

  $A_n$ is generated from  $Y_n$ by the use of only
 a finite set  $T_\varphi$ of terms of less than $k_\varphi$ variables. If
$n$ is
big enough with regard to  $k_\varphi$ and  card($T_\varphi$),  because
$q_n>2^n$, there exist parameters  $a_1, \ldots , a_k$, elements of
$P_0^{M_n}$,
and some  indiscernables $x_1, \ldots , x_j$, and  $y_1, \ldots , y_j$
of  $X_n$, such that $x_1<\ldots <x_j$
 and $y_1<\ldots <y_j$ and  $\bar x \neq \bar y$  and a term  $t\in
T_\varphi$ such that
$t(a_1,\ldots ,a_k, x_1,\ldots , x_j)<t(a_1,\ldots , a_k, y_1,\ldots , y_j)$
and these two elements  being in
$P_1^{M_n}$.

 But then   we could find in $X_n$ a sequence $(\bar {x_i})_{1\leq i\leq
N}$,
with  $N$ arbitrarily large, such that for each $i$, $1\leq i\leq N$,
$\bar x_i \bar x_{i+1}$ is of the order type
 of  $\bar x\bar y$.

 Then  for all integers $i$ such that
$1\leq i\leq N$, $P_1^{M_n}( t(a_1,\ldots , a_k, \bar x_i) )$
and the  terms $t(a_1,\ldots , a_k, \bar x_i)$ are  distinct two by two.
This would imply that,
for all integers $N\geq 1$, card$(P_1^{M_n}) \geq N$.   So there would be a
contradiction with
card$(P_1^{M_n})=p_n$ and we have proved that $L$ does not belong to $I$.

 Thus we can infer Theorem \ref{theclo} from Claim \ref{cl3}.
 The non closure under complementation of the class $LOC_\om$ can be deduced
from the
non closure under intersection and the fact that $LOC_\om$ is closed under
union (see next Theorem) or from an example, deduced from preceding proof,
of a
local \ol~ which complement is not local
 (see next remark).\ep

\begin{Rem} {\rm  The above proof shows in particular that the  \ol~
$A=\{0^n1^p2^\omega \mid  p>2^n\}$ is not  local. From which  we can easily
deduce that the
  local \ol~ $\{0^n1^p2^\omega  \mid   p\leq 2^n\}=L$ has a complement which
is not a local \ol.
(This \ol~ $L$ is local because $\{0^n1^p \mid   p\leq 2^n\}$ is a local
finitary language
\cite{loc}).
\nl Indeed if its complement was $\omega$-local, we would deduce, from a
local sentence
 $\varphi$ such that $L_\omega(\varphi)=L^-$, another local sentence
  $\psi$ such that  $L_\omega(\psi)=A$.
\nl For example the sentence $\psi$, conjunction of}:

\begin{itemize}
\ite $\varphi$ ,

\ite $\forall x y [ (P_0 (x) \wedge  P_1(y) ) \rightarrow x<y ]$ ,

\ite $\forall x y [ (P_1(x) \wedge  P_2(y) ) \rightarrow x<y ]$ ,

\ite $\forall x y [ (P_0(x) \wedge  P_2(y) ) \rightarrow x<y ]$ ,

\ite $ P_2(c)$,
where $ c$ is a new constant symbol.
\end{itemize}
\end{Rem}

\noi Now we establish that $LOC_\om$ is closed under several operations.

\begin{The}\label{theclo2} The class $LOC_\om$ is closed under union, left
concatenation
with local
(finitary) languages, $\lambda$-free substitution of local (finitary)
languages,
 $\lambda$-free morphism.
\end{The}

\proo
\nl {\bf Closure under union.}
\nl Let $\vp_1$ and $\vp_2$ be two local sentences defining
 local \ol s
$L_\om(\varphi_1)$ and $L_\om(\varphi_2)$ over a finite alphabet $\Si$.
Let us define a new local sentence
$\varphi_1 \cup \varphi_2$ which defines the local \ol~
$L_\om(\varphi_1 \cup \varphi_2)=L_\om(\varphi_1) \cup L_\om(\varphi_2)$:

 We may assume that $S(\vp_1) \cap S(\vp_2) = \Lambda_\Si$.
 Then $S(\varphi_1 \cup \varphi_2)$ will be $S(\vp_1) \cup S(\vp_2)$.
And the sentence $\varphi_1 \cup \varphi_2$ is the following sentence:

 $ [ \vp_1 \bigwedge_{{\rm n-ary ~function~ symbol}~ f \in S(\vp_2)}
( \fa x_1, \ldots , x_n f(x_1, \ldots , x_n)=min(x_1, \ldots , x_n) )]$
\nl $\bigvee [ \vp_2 \bigwedge_{{\rm n-ary~ function ~symbol}~ g \in
S(\vp_2)}
( \fa x_1, \ldots , x_n g(x_1, \ldots , x_n)=min(x_1, \ldots , x_n) )]$

\hs {\bf Closure under left concatenation by a local finitary language.}
\nl Consider a finitary local language $\Lp$ and a local \ol~ $L_\om(\psi)$
over the same
alphabet $\Ga$.
\nl We may easily assume that $L_\om(\vp)$ is empty, possibly adding a
constant symbol
$c$ to the signature of $\varphi$ and
adding the conjunction $\fa x [x\leq c]$ to the sentence $\varphi$ (this
means that every
model of $\vp$ has a greatest element).
\nl We may also assume that
$S(\varphi)\cap S(\psi)=\{<, (P_a)_{a\in\Ga}\} = \Lambda_\Ga$.
\nl Let then $P$ be a new unary  predicate symbol not in $S(\varphi)\cup
S(\psi)$, and let
$\vp.\psi$ be the following sentence in the signature $S(\varphi)\cup
S(\psi)\cup \{P\}$,
which is the  conjunction of:

\begin{itemize}
\ite  $ (<$ is a linear order ),
\ite  ($(P_a)_{a\in\Ga }$  form a partition),
\ite  $\fa xy  [ P(x) \wedge \neg P(y) \ra x<y ]$,
\ite  $\fa x_1, \ldots , x_j \in P [\vp_0 (x_1, \ldots ,  x_j) ]$,
\nl where $\vp = \fa x_1, \ldots , x_j \vp_0 (x_1, \ldots ,  x_j)$
with $\vp_0$ an open formula,
\ite  $\fa x_1, \ldots , x_m \in P [ f(x_1, \ldots ,  x_m) \in P ]$,
\nl for each m-ary function $f$ of  $S(\vp)$,
\ite  $\fa x_1, \ldots , x_m
[ \bigvee_{1\leq i \leq m} \neg P(x_i) \ra f(x_1, \ldots ,  x_m)= min(x_1,
\ldots , x_m) ]$,
\nl for each m-ary function $f$ of  $S(\vp)$,
\ite $P(c)$, for each constant $c$ of $S(\vp)$,
\ite  $\fa x_1, \ldots , x_j \in \neg P [\psi_0 (x_1, \ldots ,  x_j) ]$,
 \nl where $\psi = \fa x_1, \ldots , x_j \psi_0 (x_1, \ldots ,  x_j)$ with
$\psi_0$ an open formula,
\ite  $\fa x_1, \ldots , x_m \in \neg P [ f(x_1, \ldots , x_m) \in \neg
P ]$,
\nl for each m-ary function $f$
of  $S(\psi)$,
\ite  $\fa x_1, \ldots , x_m
[ \bigvee_{1\leq i \leq m} \ P(x_i) \ra f(x_1, \ldots ,  x_m)= min(x_1,
\ldots , x_m) ]$,
\nl for each m-ary function $f$ of  $S(\psi)$,
\ite $\neg P(c)$, for each constant $c$ of $S(\psi)$,
\end{itemize}

\noi This sentence $\vp.\psi$ is equivalent to a universal formula and
closure in its models
takes
at most $max(n_\vp, n_\psi)$ steps, hence it is a local sentence and by
construction
it holds that $L(\vp.\psi)=L(\vp).L(\psi)$. Moreover  when $\om$-words are
considered $L_\om(\vp.\psi)=L(\vp).L_\om(\psi)$ holds because by hypothesis
$L_\om(\vp)$ is
empty.

\hs {\bf Closure under $\lambda$-free substitution of local \la s.}
\nl The proof is very similar to our proof of the closure of the class $LOC$
under substitution
by local finitary languages  in \cite{loc}. We recall it now.

  Recall first the notion of substitution:
\nl  A substitution $f$ is defined by a mapping
$\Si\ra P(\Ga^\star)$, where $\Si =\{a_1,...,a_n\}$  and $\Ga$ are two
finite alphabets,
$f: a_i \ra L_i$ where $\fa i\in [1;n]$, $L_i$ is a finitary language over
the alphabet $\Ga$.
The substitution is said to be $\lambda$-free if $\fa i\in [1;n]$, $L_i$
does not contain
the empty word $\lambda$. It is a ($\lambda$-free) morphism when every
language $L_i$
contains only one (nonempty) word.
\nl Now this mapping is extended in the usual manner to finite words
and to finitary languages: for some letters $x(0)$, \ldots , $x(n)$  in
$\Si$,
 $f(x(0)x(1) \ldots  x(n))= \{u_0u_1 \ldots  u_n \mid \fa i\in [0;n]~~u_i\in
f(x(i))\}$,
 and for  $L\subseteq \Sis$,    $f(L)=\cup_{x\in L} f(x)$.
\nl If the substitution $f$ is $\lambda$-free, we can extend this to
$\om$-words and
\ol s: $f(x(0)x(1) \ldots x(n)\ldots )= \{u_0u_1 \ldots u_n \ldots  \mid \fa
i\geq 0 ~~
u_i\in f(x(i)) \}$
 and for $L\subseteq \Si^\om$, $f(L)=\cup_{x\in L} f(x)$.

  Let then $\Si =\{a_1, \ldots  ,a_n\}$ be a finite alphabet and let $f$
be a $\lambda$-free
substitution of local languages:
$\Si\ra P(\Ga^\star)$, $a_i \ra L_i$ where $\fa i\in [1;n]$,
 $L_i$ is a \loc~defined by the sentence $\varphi_i$, over the alphabet
$\Ga$.
 We may assume  that $L_\om(\varphi_i)$ is
empty, possibly adding a constant symbol $c_i$ to the signature of
$\varphi_i$ and
adding the conjunction $\fa x [x\leq c_i]$ to the sentence $\varphi_i$ (this
means that every
model of $\vp_i$ has a greatest element). We also assume that the signatures
of the
sentences $\varphi_i$ verify $S(\varphi_i)\cap S(\varphi_j)=\{<, (P_a)_{a\in
\Ga}\}$ for
$i\neq j$. Let now $L\subseteq\Si^\om$ be a local \ol~ defined by a \locs
$\varphi$.
We shall denote
$Q_{a_i}$ the unary predicate of $S(\varphi)$ which indicates the places of
the letters $a_i$
in a word of $L$, so  if $a_i\in \Ga\cap\Si$ for some indice $i$,
 there will be two distinct predicates $Q_{a_i}$  and $P_{a_i}$. We may also
assume that
$\fa i\in [1,\ldots ,n]$, $S(\varphi_i)\cap S(\varphi)=\{<\}$. Then we can
 construct a \locs $\psi$
(already given in \cite{loc}) such that  $L_\om(\psi)=f(L)$.
\nl $\psi$ is the conjunction of the following sentences, which meaning is
explained below:

\begin{itemize}
\ite $ ``<$ is a linear order '',
\ite $\forall xy [(I(y)\leq y) \wedge (y\leq x\rightarrow I(y)\leq
I(x))\wedge
(I(y)\leq x\leq y\rightarrow I(x)=I(y))]$,

\ite $ \forall x [ I(x)=x \leftrightarrow P(x)]$,

\ite $ P(c)$, for each constant $c$ of $S(\varphi)$,

\ite $ \forall x_1, \ldots , x_k [R(x_1, \ldots , x_k) \rightarrow
P(x_1)\wedge \ldots
\wedge P(x_k)]$,
 \nl for each predicate $R(x_1, \ldots , x_k)$ of  $S(\varphi)$,

\ite $ \forall x_1, \ldots , x_j [(P(x_1)\wedge \ldots \wedge P(x_j))
\rightarrow
P(f(x_1, \ldots , x_j)) ]$,
\nl for each j-ary function symbol f of $S(\varphi)$,

\ite $ \forall x_1, \ldots , x_j [ ( \bigvee_{1\leq i \leq j} \neg P(x_i))
\rightarrow
 f(x_1, \ldots , x_j)=min(x_1, \ldots , x_j) ]$,
\nl for each j-ary function symbol f of
 $S(\varphi)$,

\ite $ \forall x_1, \ldots , x_m [ (P(x_1) \wedge \ldots \wedge P(x_m))
\rightarrow \varphi_0
 (x_1, \ldots , x_m) ]$,
\nl where
$\varphi = \forall x_1, \ldots , x_m \varphi_0(x_1, \ldots , x_m)$
 with  $\varphi_0$ an open formula,

\ite $ \forall x_1, \ldots , x_j [\bigvee _{i,k \leq j}(I(x_i)\neq I(x_k))
 \rightarrow f(x_1, \ldots , x_j)= min(x_1, \ldots , x_j) ]$,
\nl for every function f of
 $S(\varphi _l)$ for  an integer $l\leq n$,

\ite $ \forall xy_1, \ldots , y_j [ (\bigwedge _{1\leq l\leq j} I(y_l)=I(x))
\rightarrow
 I(f(y_1, \ldots , y_j))=I(x) ]$,
\nl for each j-ary function symbol f of  $S(\varphi_i)$ for
 an integer $i\leq n$,

\hspace{12mm}   Finally, for each $i\leq n$:

\ite $  \forall x y_1, \ldots , y_p
[ (\bigwedge _{1\leq l \leq p}I(y_l)=I(x) \wedge Q_{a_i}(I(x))  )
 \rightarrow \varphi_i^0 (y_1, \ldots , y_p) \wedge \{ (e_j(y_1)=I(x) \wedge
f_j(y_1, \ldots , y_p)=y_1 \wedge
\neg R_j(y_1, \ldots , y_p) \wedge \bigwedge_{e_i \in S(\varphi_i)}
e_i(y_1)=e_i(x);$
where  $n\geq j\neq i$,
 and  $e_j, f_j, R_j$ run over the  constants, functions,
and  predicates of $S(\varphi_j) \} ]$,

\end{itemize}

 \noi Above , 1) to each constant $e_l$ of  $S(\varphi_l)$ is associated a
new unary function
 $e_l(y)$ and  2) whenever $\varphi_i =\forall y_1, \ldots , y_p \psi_i
(y_1, \ldots , y_p) $
with $\psi_i$
an  open formula, $\varphi_i^0$ is  $\psi_i$ in which every constant $e_i$
has been replaced
 by the function $e_i(y)$.

 {\bf Construction of  $\psi$ :}
\nl   Using the function  $I$ which  marks  the first letters of the
subwords,
 we divide an $\om$-word into omega (finite) subwords (the function $I$ is
constant on
each subword and $I(x)$ is the first letter of the subword containing $x$).
In every model  $M$ of order type $\omega$ of $\psi$,
 the set of the ``first letters of subwords" , $P^M$, grows richer
in a model of order type $\omega$
of $\varphi$ (therefore will constitute an $\om$-word of $L$).
\nl Then we ``substitute": for each letter $a_i$ in $P^M$,
 we substitute a (finite) word of $L_i$, using for that the formula
 $\varphi_i$.
\nl If closure takes at most $n_\varphi$ (respectively $n_{\vp_i}$) steps in
every
model of $\vp$ (respectively of $\vp_i$), then closure takes at most
 $[1 + n_\varphi +  sup_i (n_{\vp_i})]$ steps in each model of $\psi$ (one
takes closure under
the function $I$, then under the functions of $S(\vp)$, and finally under
the functions of
$S(\vp_i)$, $1\leq i\leq n$).
\nl Therefore $\psi$ is a \locs and by construction $\psi$ defines the
$\om$-language $f(L)$.

\hs {\bf Closure under $\lambda$-free morphism.}
\nl It is just a particular case of the preceding one, when every language
$L_i$ contains only
one non-empty finite word. \ep

\hs {\bf Acknowledgments.}
 Thanks  to the anonymous referees  for useful comments on the
 preliminary version of this paper.

\begin{footnotesize}

\end{footnotesize}

\end{document}